\documentclass[prl,
showpacs,
preprintnumbers,
amsfonts,
amsmath,
amssymb,
nobalancelastpage,
superscriptaddress
,reprint
]{revtex4-1}

\usepackage{graphicx}
\usepackage{color}
\usepackage{epsfig}
\usepackage{dcolumn}
\usepackage{bm}
\usepackage[latin2]{inputenc}

\begin{document}

\author{L. Czekaj}
\affiliation{Faculty of Applied Physics and Mathematics,
Gda{\'n}sk University of Technology, 80-952 Gda{\'n}sk, Poland}
\affiliation{National Quantum Information Center of Gda\'nsk, 81-824 Sopot, Poland}

\author{R. W. Chhajlany}

\affiliation{Faculty of Physics, Adam Mickiewicz University,
  Umultowska 85, 61-614 Pozna\'{n}, Poland}

\affiliation{Faculty of Applied Physics and Mathematics, Gda{\'n}sk
  University of Technology, 80-952 Gda{\'n}sk, Poland}

\affiliation{National Quantum Information Center of Gda\'nsk, 81-824
  Sopot, Poland}

\author{P. Horodecki}
\affiliation{Faculty of Applied Physics and Mathematics, Gda{\'n}sk
  University of Technology, 80-952 Gda{\'n}sk, Poland}

\affiliation{National Quantum Information Center of Gda\'nsk, 81-824
  Sopot, Poland}

\date{\today}
\title{Directed percolation effects emerging from superadditivity of
  quantum networks}

\begin{abstract}
  Entanglement indcued non--additivity of classical communication
  capacity in networks consisting of quantum channels is considered.
  Communication lattices consisiting of butterfly-type entanglement
  breaking channels augmented, with some probability, by identity
  channels are analyzed.  The capacity superadditivity in the network
  is manifested in directed correlated bond percolation which we
  consider in two flavours: simply directed and randomly oriented. The
  obtained percolation properties show that high capacity information
  transfer sets in much faster in the regime of superadditive
  communication capacity than otherwise possible.  As a
byproduct, this sheds light on a new type of entanglement based
quantum capacity percolation phenomenon.

\end{abstract}

\maketitle

{\it Introduction.} Percolation (see {\it e.g.} \cite{Stauffer-book,
  Grimmett-book}) is a natural concept that emerges in the description
of spreading processes in the presence of medium imperfections.
Percolation effects in quantum networks have recently been the subject
of increasing interest
\cite{AcinEtAl2008,Wehr,PerseguersEtAl2008,BraodfootEtAl2009,Perseguers2010}. Most
of the attention has been restricted to the generation of large scale
networks with maximally entangled states between elementary nodes to
allow for quantum communication applications, starting from initial
imperfect, {\it i.e.}  non-maximally entangled state networks. The
interesting central new insight introduced in \cite{AcinEtAl2008,
  Wehr} is that local quantum operations may be used not only to
purify entanglement but, simultaneously, to change the topology of the
lattice to a new one with lower percolation probability threshold.
This idea has been developed for different states
\cite{PerseguersEtAl2008,BraodfootEtAl2009} and lattice dimensions
\cite{Perseguers2010,GrudkaEtAl2011}. In a
different context, percolation concepts also appeared in quantum
information theory (QIT) in the study of  cluster state
generation \cite{KielingEisert2009}.

The capacity of a network determines its utility in the domain of
communication. The development of QIT has led to the uncovering of
interesting quantum effects on channel capacities, {\it e.g.}
superadditivity of quantum (Q-type) channel capacity
\cite{SmithYard2008}. This result followed the intuition developed in
the bound entanglement activation effect \cite{Activation} (where two
weak resources activate each other becoming collectively useful for
some task) continued further in Refs. \cite{Superactivation} and
\cite{DuerEtAlChannels}. Independently, the first superadditivity
effect of classical (C-type) capacity in quantum multi-access channels
has been described \cite{CzekajHorodecki2009}. Both Q-type and C-type
superadditivities have been proven even for entanglement breaking
channels \cite{GrudkaHorodecki2010}.

In this paper, we consider percolation effects in quantum networks
from a channel perspective. In particular, we show that channel
superadditivity can be used to enhance percolation of information
through networks. The schemes are based on network models of classical
information transfer through quantum multipartite channels (MACs),
which can be mapped to certain types of {\it directed} bond
percolation problems \cite{Broadbent}. We first consider a simple
layered communication scheme (A) to demonstrate the basic idea of
percolation assisted by superadditive capacities and then describe a
more complicated scheme of multidirectional communication
(B). Interestingly, the percolation problem in case B does not seem to
have been studied elsewhere in the literature.

The basic ingredients of the quantum networks are: a {\it passive}
fixed underlying network built up of elementary entanglement breaking
MACs, and an {\it active} auxiliary incomplete network consisting of
randomly generated (open) bonds.  The term passive means that no bond
in the network allows {\it a-priori} high capacity communication
(HCC), whereas elements of the active network are high capacity
channels. Importantly, the active channels can serve to activate HCC
through the passive channels.

 \begin{figure}[t]
  	\centering
    \includegraphics[scale=0.31]
    {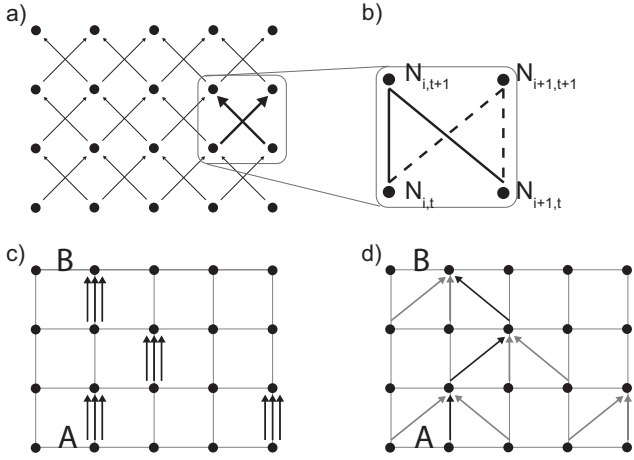}
    \caption{\label{fig:model_A} Model A - layered communication
      network: a) passive network; users are located at nodes of a
      square lattice, the lattice is filled with butterfly shaped
      primitives, information flow is directed from layer
      $t\rightarrow t+1$; b)  butterfly primitive consists of two
      MACs (solid and dashed wedges) from
      Ref.~\cite{GrudkaHorodecki2010}; Node (user) $N_{i,t}$ can communicate
      with $N_{i-1,t+1},N_{i,t+1},N_{i+1,t+1}$ with maximal capacity
      $C'\ll C_{0}$; c) active network filled randomly by triples of ideal
      channels allowing HCC (no HCC from A to B); d)  using entanglement changes the
      geometry of the HCC network leading to directed communication
      paths (black arrows) with capacities $\geq C_0$ from A to B.  }
  \end{figure}

 \begin{figure}[t]
  	\centering
      \includegraphics[scale=0.3]
     {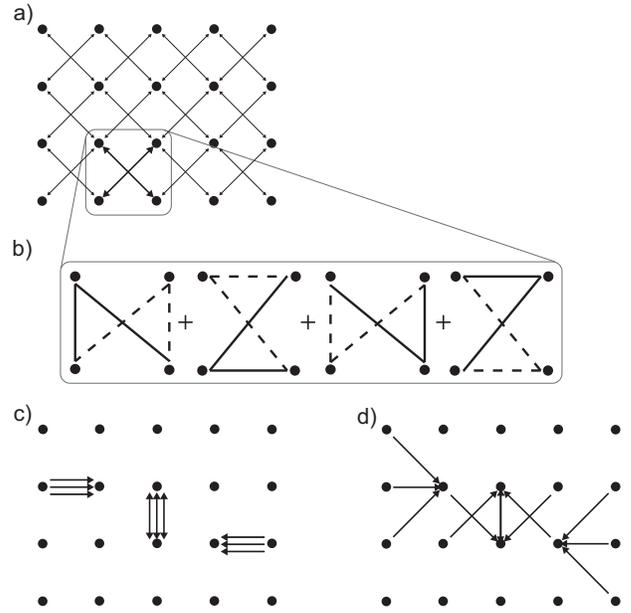}
     \caption{Model B - Multidirectional communication networks: a)
       passive network, allowing communication in each direction; b)
       structure of butterfly primitive; c) active channels; d)
       emerging network geometry induced by entanglement between
       active and passive
       channels.  \label{fig:rys_3_model_B+activation}}
  \end{figure}

  {\it Model A - Layered network communication.} The elementary
  channel of the passive network is chosen in this paper as a 2-sender
  1-receiver {\it noisy} quantum MAC depicted as the wedge shaped
  channels in Fig. \ref{fig:model_A}(b): the slanted line is a
  two-qudit ($d^{2}$--dimensional) input pertaining to one user; the
  vertical line is a single qudit input of the second user. The single
  qudit system is modified by one of a chosen set of (orthogonal)
  $d^{2}$ unitary operations fired by the logical value of the
  two-qudit system (see Ref. \cite{GrudkaHorodecki2010} for
  details). This single qudit line is further modified by a
  depolarizing channel and is the sole output system at the receiver's
  end. For moderate to large depolarization -- the working regime
  considered here -- this channel has poor classical capacity $C'$ for
  any user \cite{GrudkaHorodecki2010}:
  \begin{gather}
C' \ll C_{0} \leq C_{{\rm max}} \equiv \log_{2}d,
\label{Rpab}
\end{gather}
where $C_{max}$ is the maximal attainable capacity for a single qudit
output, corresponding to an identity channel. Communication can be
improved by adding a high capacity active channel, {\it e.g.} an ideal
channel, along the vertical transmission line.  Inputting a two-party
maximally entangled state to the ideal channel and vertical line of a
wedge channel increases the capacity along the slanted line of the
wedge channel to a much higher value $C_{0}$ (see Eq.(\ref{Rpab})) --
a manifestation of superadditivity of channel capacities
\cite{comment:1} (for high depolarization when the channel becomes
entanglement breaking, $C_{0} \approx (d+1)C'$
\cite{GrudkaHorodecki2010}). We define HCC as transmission at a rate
$\geq C_{0}$.

The structure of the wedge channel induces a sense of direction of
communication. In the network context, we first consider the scenario
where only forward communication is allowed (see
Fig.\ref{fig:model_A}).  The passive channel inputs are shared by
nearest neighbour pairs of sites in horizontal layers giving rise to a
butterfly shaped fixed network.  The active channels are only placed
on vertical bonds. Suppose that these are available with probability
$p$ -- one can assume that identity channels are initially available
at all vertical bonds, yet due to fragility w.r.t. noise either
remain useful (ideal) with probability $p$ or become unuseful random
operations with probability $1-p$ effectively erasing information.  In
our scheme, we consider a setup where a triple of identity channels is
the basic active channel \cite{comment:2}-- this allows the
possibility of simultaneous HCC between a given sender and his
vertically placed receiver as well as between the sender's two
horizontal nearest neighbours and that receiver
(Fig. \ref{fig:model_A}).

The question of establishing long range HCC in such a network is a
directed percolation problem. One asks, under what conditions, is it
possible for any user to be able to perform directed HCC, through
intermediate nodes, with a user or users located at a distance scaling
with the length of the network.  One may compare two scenarios - (a)
entanglement-free or classical, where no entanglement is allowed in
the protocol used at any node and (b) entanglement-assisted (EA),
which takes full advantage of the superadditive effect described
earlier.  The former case corresponds to communication only along
1--dimensional paths and the percolation threshold probability is $1$,
rendering the network useless for HCC for finite loss probability of
active channels. The EA scheme involves changing the geometry of the
HCC network and is mapped to a {\it correlated} directed bond
percolation problem where with probability $p$ three directed bonds
(forming an arrow shape) are placed on the lattice
(Fig.\ref{fig:model_A}(e)). The threshold probability is significantly
suppressed due to entanglement induced increased connectivity.  We
performed a standard Monte Carlo simulation (see \cite{supplement})
and found the percolation threshold to be $p_{c}=0.5388$ with accuracy
$\Delta =0.0005$.  We have checked that the studied percolation
transition lies, as expected, in the Directed Percolation (DP)
universality class, by computing a complete set of critical exponents
and found them to be in agreement with those obtained for the
uncorrelated directed bond percolation problem.  The important basic
characteristic of directed percolation is that the connected clusters
of nodes are geometrically highly asymmetrical and are restricted to
acute cones (with cone angle $\pi /2$ when $p=1$) with axes along the
vertical lines passing through the starting nodes.  Near the critical
point the clusters are very narrow and essentially quasi
1-dimensional, as the probability of obtaining a connection with a
site a large distance away from the source and at an angle $\theta $
from the axis, $\Theta (p, \theta ) >0$ for $|\theta | <\delta \theta
(p) \sim (p-p_{c})^{b}$ (see {\it e.g.} Dhar and Barma
\cite{DharBarma1980}).  The EA scheme beats the classical scheme in
that there is a wide window in $p$ for HCC, and that the horizontal
extent of connections $|\theta |$ at a distance $t \rightarrow \infty
$ away changes from $0$ to $\pi /4$ as $p$ increases from $p_{c}$ to
1.

  \begin{figure}[t]
  	\centering
  		\includegraphics[scale=0.475]{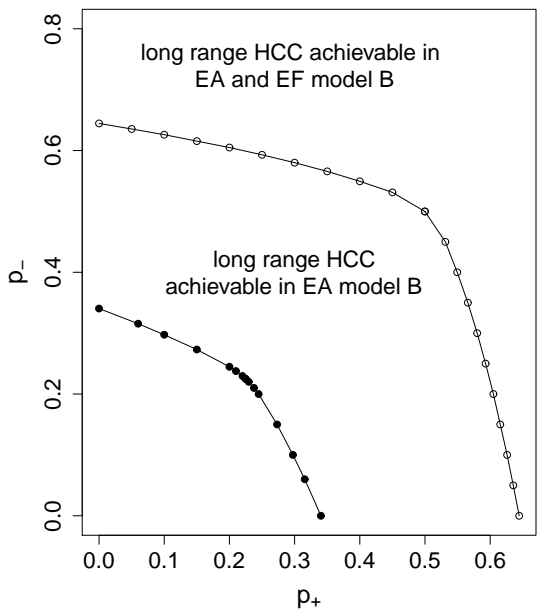}
        \caption{(a) Comparison of the percolation critical lines
          $(p_{+}^c,p_{-}^c(p_{+}^c))$ for percolation process on
          randomly oriented square lattice (classical scheme) and
          butterfly network (EA scheme).  Dots are numerically
          obtained points.
          }
\label{fig:percTresModelB-3.0}
  \end{figure}

  {\it Model B -- Randomly oriented communication.} We now move to a
  general scenario within the described framework of channels
  (Fig. \ref{fig:rys_3_model_B+activation}). Now each node can
  communicate in {\it both} directions with its nearest neighbours
  (main bonds) on a square lattice as well as all of its next nearest
  neighbours (diagonal bonds) through the passive network. The
  configuration leading to this possibility is shown in
  Fig.\ref{fig:rys_3_model_B+activation} (a) and (b). The active
  network channels are placed along the main bonds of the square
  lattice and are tagged for use in a particular direction.

   \begin{figure}[t]
   	\centering
      \includegraphics[scale=0.475]{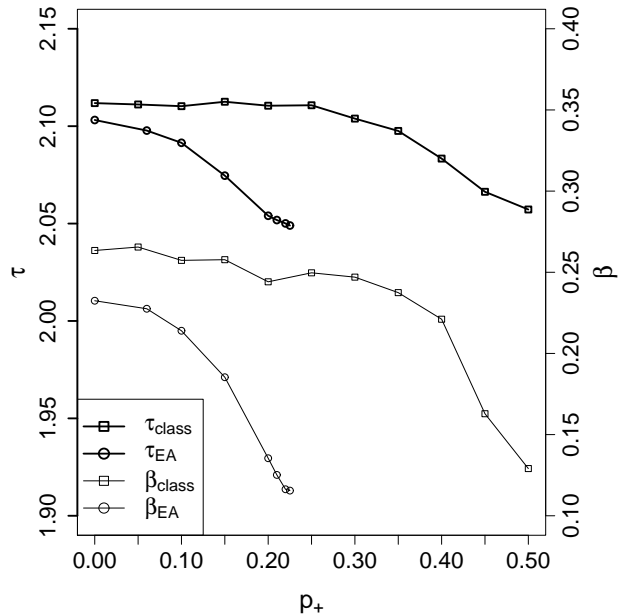}
      \caption{Comparison of critical exponents $\tau $, $\beta $ for
        classical and EA schemes of multidirectional oriented
        percolation.}
\label{fig4}
   \end{figure}

   HCC between distant nodes is now a multidirectional bond
   percolation problem. We consider the case where upward,
   left-to-right oriented identities are present with probability
   $p_+$ while downward, right-to-left oriented identities are placed
   with probability $p_-$. Note, that bidirectional bonds appear as
   the independent choice of two oppositely directed channels on a
   given bond.

   This setup is interesting already for the case when, say,
   $p_{+}=0$. Using an entanglement free protocol, capacity $C \geq
   C_{0}$ can already be obtained through the main bonds of the square
   lattice along the up or right directions, in contrast to Model
   A. This is the well known square lattice directed bond percolation
   problem for which the percolation threshold $p_{c} \approx 0.64$
   (see {\it e.g.}  \cite{DharBarma1980}). But, in the EA scheme, due
   to increased and correlated connectivity, the percolation threshold
   is almost halved w.r.t. the classical scheme and has been
   calculated here to be $p_{c} \approx 0.34$. Both phase transitions
   lie in the DP universality class. At moderate $p \approx 0.6$, the
   probability of a starting node belonging to an infinite or system
   spanning anisotropic cluster (in the thermodynamic limit)
   $F_{\infty } > 1/2 $ using superadditivity effects while $F_{\infty
   }=0$ otherwise.

   The general multidirectional communication problem is the most
   interesting setup. In the classical scenario, this reduces to
   square lattice randomly oriented percolation. Problems of this
   category were studied in the context of random resistor diode
   networks (see {\it e.g.}  \cite{Redner}) mainly using
   renormalization group calculations. However not many results seem
   to be available in the literature on the $p_{\pm}$ percolation
   problem (see however \cite{grimmett} for some analytical
   properties).

   Here, we map out the phase diagram in the $(p_{+}, p_{-})$ plane in
   Fig.\ref{fig:percTresModelB-3.0} using Monte Carlo simulations (see
   Appendix).  Note that the diagram is symmetrical w.r.t. the line
   $p_{+}=p_{-}$ due to system symmetry under the interchange $p_{+}
   \leftrightarrow p_{-}$. This result is to be compared with the
   multidirectional correlated bond percolation phase diagram for the
   EA scheme (Fig. \ref{fig:rys_3_model_B+activation}(d)). As can be
   seen in Fig.\ref{fig:percTresModelB-3.0}, the EA scheme is
   drastically better than the classical scheme in the whole parameter
   plane.

   An interesting feature of both schemes is that they allow switching
   of universality classes of phase transitions from the DP class to
   the Isotropic Percolation (IP) class to which standard bond/site
   percolation belongs. This is facilitated by the choice of
   parameters $p_{\pm}$, and as a result one may accordingly change
   the properties of the long range clusters. To provide evidence for
   this, we calculate two universal critical exponents
   (Fig. \ref{fig4}) $\beta $ and the Fisher exponent $\tau $ (see
   \cite{supplement}) as one moves along the critical lines
   (Fig. \ref{fig:percTresModelB-3.0}) of the two models. Recall, that
   $\beta $ determines how the percolation probability $F_{\infty }
   \sim (p-p_{c})^{\beta }$ increases above the percolation threshold,
   while the Fisher exponent determines how the probability of
   obtaining a cluster larger than size $n$ decreases with $n$ at the
   critical point $p=p_{c}$: $F_{n} \sim 1/n^{\tau -2}$.  Note that
   the values of these exponents in the DP class are $\beta \approx
   0.276, \tau \approx 2.112$ \cite{DharBarma1980} and in IP: $\beta =
   5/36 \approx 0.139, \tau = 187/91 \approx 2.0549$
   \cite{Stauffer-book}. First consider square lattice
   multidirectional percolation in Fig.\ref{fig4} (the classical
   scheme). Both exponents show that as one moves along the critical
   line, the phase transitions lie in the DP universality class until
   $p_{-}^{c} \approx 0.4$, since this region inherits the exponents
   at the DP point $p_{-}^{c}=0$. At $p_{-}^{c} =1/2 (= p_{+}^{c})$,
   which is an isotropic symmetry point, one obtains values of the
   exponents corresponding to the IP universality class. In between,
   there is a characteristic crossover region between the two types of
   behaviour.  In particular, this means that for a choice of
   parameters approaching the isotropic point, one obtains different
   characteristic growth of clusters (determined by different critical
   exponents) than when in the DP region. Secondly the cluster
   geometrical characteristics change from highly anisotropic to
   isotropic. The butterfly network percolation problem (EA scheme)
   basically follows the same pattern (see Fig. \ref{fig4}) and can be
   considered a rescaled version of the classical problem, wherein
   again lies the quantum advantage of the EA scheme. For this
   network, the isotropy point is found to be located at
   $p_{+}^{c}=p_{-}^{c} \approx 0.225$.  The slightly lower values of
   $\beta $ in the DP regime as compared to the square lattice problem
   do not seem to be significant - and are a result of the extreme
   sensitivity of the calculated exponents on the accuracy of critical
   probabilities.

   {\it Concluding remarks.}  We have described percolation effects in
   a channel context showing how directed percolation effects (not
   considered before) emerge in the consideration of quantum networks.
   In the context of percolation theory, to our knowledge, the
   multidirectional correlated bond percolation problem has not been
   studied before. Finally, quite remarkably, our results provide a
   new entanglement percolation effect in the spirit of
   Ref. \cite{AcinEtAl2008}. Keeping everything else unchanged,
   consider the Bell measurement (BM) channel of Fig. 1 of Ref.
   \cite{GrudkaHorodecki2010} instead of the MAC used here, with AC
   (BC) playing the role of diagonal (resp. vertical) bond of the
   square lattice.  Then any randomly generated singlet between B and
   C can be switched, via entanglement swapping, to a singlet on the
   diagonal AC. This leads to the same geometry as discussed here but
   now one asks about the possibility of building a long range network
   of singlets. Note that singlets are directionless. Without using
   the BM channel, we obtain a ``classical'' scheme related to square
   lattice percolation which has threshold $p_{c}=0.5$. Since
   directionless percolation must certainly be at least as good as
   directed percolation, an EA scheme making use of the switching
   mechanism of the BM channel will have threshold at most equal to
   the calculated threshold at the isotropic point of Model B
   $p_{c}^{\text EA}\leq p_{+}^{c}=p_{-}^{c}\approx 0.225$.

%  Consequently, it is not difficult to see, that
%    replacing the two-user primitive in Fig. \ref{fig:model_A} b) of
%    the present paper by the ABC one discussed above leads to virtually
%    the same geometry as in Fig. 2 with one-side directed arrows
%    replaced by bonds formed by d-singlets (two-side directed arrows
%    give pairs of d-singlets).  Focusing now on entanglement
%    percolation we see that for $p_{+}= p_{-}$ the bond percolation is
%    known to be $0.5$ while the EA percolation region is definitely
%    better than that of Fig. 3 (since at the moment we ask about
%    isotropic i.e. not directed percolation). This gives the
%    improvement of entangled-bonds percolation form $p_{c}= 0.5$ to
%    $p_{c}\leq 0.22$.

 \begin{acknowledgments}
   The authors thank J.K. Korbicz for discussions. 
   P. H. thanks also C. H. Bennett and A. Grudka for discussions.  The work was supported
   by EU project QESSENCE and by the Polish Ministry
   of Science and Higher Education through Grant No. NN202231937.
   Part of the work was done in Quantum Information Centre of Gdansk.
 \end{acknowledgments}

\cleardoublepage
\appendix
\begin{widetext}

{\noindent \large{Supplemental Material}}

\begin{center} {\large \bf Directed percolation effects emerging from
    superadditivity of quantum networks}
 \smallskip

{\L}ukasz Czekaj, Ravindra W. Chhajlany and Pawe{\l} Horodecki

\smallskip

\end{center}

 \setcounter{page}{1}
  \pagenumbering{roman}

\section{Methods }\label{}

The critical percolation properties of the two models were studied
using direct Monte Carlo simulations in the spirit of Dhar and Barma
\cite{DharBarma1980a}.  We studied the change in behaviour of the
probability $F_{n}$, of appearance of clusters of size greater than
$n$, with bond probabilities $p$ and looked for characteristic scaling
behaviour expected in the critical region to identify percolation
thresholds plotted. Below the probability threshold, $F_{n} \propto
\exp(-n)$ for large $n$ while $F_{n} \rightarrow {\rm constant}$ in
the supercritical phase \cite{Stauffer-book}. The critical region is
characterized by scaling laws with $F_{n} \propto n^{2-\tau }$
described by a power law dependence at the critical value $p_{c}$,
which we localized by sweeping through super-- and sub--critical
probabilities using an interval bisection method. For model A, cluster
size distribution data was obtained by performing $10^{5}$
realizations (per value of probability $p$) of cluster growth starting
from a single node.  Model $B$ simulations were performed on a fixed
$2\times 10^3 \text{ by } 2\times 10^3$ square lattice also with
$10^{5}$ realizations for each $p$, where cluster connectivity was
identified using a breadth--first search algorithm.

The qualitative values of the critical exponents $\tau, \beta  $ presented in the paper for
model B are defined as follows:
\[
F_{n}(p) \sim n^{-(\tau -2)} \text{ at } p=p_{c}
\label{Rzbb}
\]
and
\[
F_{\infty }(p) \sim (p-p_{c})^{\beta }
\label{Rybb}
\]
A simple method used to obtain these was to directly calculate the
slope of the plots of these two functions in log-log scale. Since the
problem contains two parameters $p_{+},p_{-}$, we chose one of them
$p_{+}$ do be the independent parameter with $p_{-}(p_{+})$ determined
so as to be on the critical line of the model. For the determination
of $\beta $, we considered clusters of size $n$ greater than $10^5$ to
be ``infinite'' or system spanning clusters on the finite lattice.

Alternatively, we also determined $\beta $ from $\tau $  and an
auxiliary exponent $\gamma  $, which describes the critical behaviour
of mean cluster size $\langle n \rangle $ also readily available from
the simulation:
\[
\langle n \rangle \sim (p_{c} -p )^{- \gamma } \text{ for } p-p_{c} \rightarrow 0^{+}
\label{R2bb}
\]
The following universal equation, derived from scaling relations, is known to
hold for directed (uncorrelated) bond percolation \cite{DharBarma1980a} and isotropic
percolation \cite{Stauffer-booka}:
\[
\beta = \Big(\frac{\tau -2}{3-\tau }\Big)\gamma
\label{R3bb}
\]
We assumed the equation to be true for the entire critical plane and
obtained results for $\beta $ in agreement with those obtained using the first
method. Results obtained in the latter manner are those presented in
the paper.

\clearpage
\end{widetext}


\begin{thebibliography}{40}

\bibitem{Stauffer-book} D. Stauffer and A. Aharony, Introduction to
  percolation theory, Taylor \& Francis 2003.

\bibitem{Grimmett-book} G. Grimmett, Percolation,
  Springer-Verlag 1999.

\bibitem{AcinEtAl2008} A. Acin, J. I. Cirac, and M. Lewenstein, Nature Phys. {\bf 3},
 256 (2007).

\bibitem{Wehr} G. J. Lapeyre, Jr., J. Wehr, and M. Lewenstein,
  Phys. Rev. A {\bf 79}, 042324 (2009).

\bibitem{PerseguersEtAl2008} S. Perseguers, L. Jiang, N. Schuch,
  F. Verstraete, M. Lukin, J. Cirac, and K. Vollbrecht, Phys. Rev. A
  {\bf 78}, 062324 (2008); S. Perseguers, D. Cavalcanti,
  G. J. Lapeyre, Jr., M. Lewenstein, and A. Acin, Phys. Rev.  {\bf A
    81}, 032327 (2010).


 \bibitem{BraodfootEtAl2009} S. Broadfoot, U. Dorner, and D. Jaksch,
   EuroPhys. Lett.  88, 50002 (2009).

 \bibitem{Perseguers2010} S. Perseguers, Phys. Rev. A 81, 012310
   (2010).


\bibitem{GrudkaEtAl2011} A. Grudka {\it et al.}, manuscript in preparation (2011).
\bibitem{KielingEisert2009}
K. Kieling, J. Eisert, ,,Percolation in quantum computation and communication'', in: {\it Quantum and Semi-classical Percolation and Breakdown
in Disordered Solids}, pages 287-319 (Springer, Berlin, 2009).


\bibitem{SmithYard2008} G. Smith, J. Yard, Science {\bf 321}, 1812 - 1815 (2008).
\bibitem{Activation} P. Horodecki, M. Horodecki, R. Horodecki, Phys. Rev. Lett. {\bf 82}, 1056 (1999).

\bibitem{Superactivation}P. Shor, J. A. Smolin, and A. V. Thapliyal, Phys. Rev. Lett. {\bf 90}, 107901 (2003).

\bibitem{DuerEtAlChannels} W. Duer, J. I. Cirac, and P. Horodecki,Phys. Rev. Lett. {\bf 93}, 020503 (2004).


\bibitem{CzekajHorodecki2009} L. Czekaj and P. Horodecki, Phys. Rev. Lett. {\bf 102}, 110505 (2009).

\bibitem{GrudkaHorodecki2010} A. Grudka and P. Horodecki, Phys. Rev. A {\bf 81}, 060305(R) (2010).

\bibitem{Broadbent}
S. R. Broadbent and J. M. Hammersley, Proc. Camb. Phil. Soc. {\bf 53}, 629 (1957).

\bibitem{DharBarma1980}D. Dhar and M. Barma, J. Phys. C: Solid State Phys., {\bf 14}, L1 (1981).

\bibitem{Hinrichsen2000} H. Hinrichsen, Adv.Phys. {\bf 49}, 815 (2000).

\bibitem{Redner} S. Redner, J. Phys. {\bf A}: Math. Gen. {\bf 14}, L349
  (1981); {\it ibid} {\bf 15}, L685 (1982); Phys. Rev.  {\bf B 25}, 3242 (1982).


\bibitem{grimmett}
Geoffrey R. Grimmett, Random Structures \& Algorithms {\bf 18} 257 (2001)


\bibitem{comment:1} This is a superadditive effect as the
  identity is unavailabe to the user of the slanted input line of
  the wedge channel.

\bibitem{comment:2} For qubit channels, {\it e.g.} this can be
  physically realized as a process of ideal transmission with
  probability $p$ of a single photon having three frequency states and
  two polarization degrees of freedom for each frequency, where the
  dominating noise process is photon loss.

\bibitem{supplement} See supplemental material for methods.

\end{thebibliography}

\begin{thebibliography}{5}
 \bibitem{DharBarma1980a}D. Dhar and M. Barma, J. Phys. C: Solid State
   Phys., {\bf 14}, L1 (1981).
\bibitem{Stauffer-booka} D. Stauffer and A. Aharony, Introduction to
  percolation theory, Taylor \& Francis 2003.
\end{thebibliography}
\end{document}